\documentclass[conference]{IEEEtran}
\IEEEoverridecommandlockouts
\usepackage{cite}
\usepackage{balance}
\usepackage{easybmat}
\usepackage{stfloats}
\usepackage{amsmath,amssymb,amsfonts}
\usepackage{graphicx}
\usepackage{textcomp}
\usepackage[linesnumbered,lined,boxed,commentsnumbered, ruled]{algorithm2e}
\usepackage{epstopdf}

\usepackage{caption}
\usepackage{subcaption}
\usepackage{algorithmic}

\usepackage[letterpaper, left=0.7in, right=0.7in, bottom=1in, top=0.7in]{geometry}

\newtheorem{thm}{Theorem}

\newtheorem{lem}{Lemma}

\DeclareFontFamily{U}{mathx}{\hyphenchar\font45}
\DeclareFontShape{U}{mathx}{m}{n}{
      <5> <6> <7> <8> <9> <10>
      <10.95> <12> <14.4> <17.28> <20.74> <24.88>
      mathx10
      }{}
\DeclareSymbolFont{mathx}{U}{mathx}{m}{n}
\DeclareFontSubstitution{U}{mathx}{m}{n}
\DeclareMathAccent{\widecheck}{0}{mathx}{"71}

\newenvironment{skproof}{\paragraph*{Sketch of Proof}}{\hfill$\square$}

\usepackage{xcolor}

\newcommand{\bs}[1]{\ensuremath{\boldsymbol{#1}}}
\newcommand{\mrm}[1]{\ensuremath{\mathrm{#1}}}
\renewcommand{\vec}[1]{\ensuremath{\mathrm{vec}}}

\newcommand{\MP}{\ensuremath{M_\mrm{P}}}

\newcommand{\diag}{\mathrm{diag}} 

\newcommand{\comment}[1]{}

\renewcommand{\baselinestretch}{.94} 
\begin{document}

\title{Decentralized Multi-Antenna Architectures with Unitary Constraints}
\author{Juan Vidal Alegr\'{i}a, Ove Edfors\\
\IEEEauthorblockA{Department of Electrical and Information Technology, Lund University, Lund, Sweden\\} 
juan.vidal\_alegria@eit.lth.se, ove.edfors@eit.lth.se}

\maketitle

\begin{abstract}
The increase in the number of base station (BS) antennas calls for efficient solutions to deal with the increased interconnection bandwidth and processing complexity of traditional centralized approaches. Decentralized approaches are thus gaining momentum, since they achieve important reductions in data/processing volume by preprocessing the received signals before forwarding them to a central node. The WAX framework offers a general description of decentralized architectures with arbitrary interplay between interconnection bandwidth and decentralized processing complexity, but the applicability of this framework has only been studied assuming unrestricted baseband processing. We consider an adaptation of the WAX framework where the decentralized processing has unitary restriction, which allows for energy-efficient implementations based on reconfigurable impedance networks at the cost of some performance loss. Moreover, we propose an effective method to minimize the performance gap with respect to centralized processing. The previous method gives a first step towards characterizing the information-lossless trade-off between interconnection bandwidth and processing complexity in decentralized architectures with unitary constraints.
\end{abstract}

\section{Introduction}
Next generation mobile broadband communication systems seem to favor the use of base station (BS) technoglogies relying on a large number of antennas due to the associated gains in spatial resolution and spectrum efficiency. This trend is embodied in massive multiple-input multiple-output (MIMO) \cite{marzetta}, which has been key towards the development of 5G \cite{emil_next}, but even more massive approaches, such as large intelligent surface (LIS)\cite{husha_data}, are being considered towards 6G. 

Increasing the number of BS antennas \cite{rusek}, although extremely beneficial for efficient usage of the frequency spectrum, comes at the cost of high interconnection bandwidth and processing complexity, specially when considering traditional centralized approaches \cite{lumami,dec_studer}. Thus, research efforts have been directed towards proposing decentralized approaches where part of the processing is distributed to reduce the volume of the data that has to be transmitted and processed at a single node \cite{dec_studer,isit_2019}. 

A novel framework, which we hereby refer to as the WAX framework, was recently introduced in \cite{wax_journal} to generalize a number of decentralized architectures with different levels of processing complexity and interconnection bandwidth, and whose information-lossless trade-off has been formally characterized.\footnote{By information-lossless we mean that there is no information loss with respect to centralized processing.} The WAX framework consists of dividing the processing into three stages, a set of linear decentralized filters applied at the antenna/panel nodes and whose size is determined by the decentralized processing complexity, a fixed combining module that merges the outputs from the decentralized filters and reduces the dimension according to the desired interconnection bandwidth, and a subsequent linear processing stage applied at the central processing unit (CPU). In \cite{A_struct}, more results were presented on how to practically implement the WAX framework by considering specific combining modules and respective decentralized schemes with generalization to any point in the mentioned trade-off. 

In \cite{wax_journal} and \cite{A_struct}, the WAX framework considered no restriction on the decentralized filters, which is reasonable if we assume that these are applied in baseband. In this work, we study a restricted version of the WAX framework where the decentralized filters, as well as the combining module, are constrained to unitary matrices. This inevitably leads to a loss in performance over the original WAX framework, but it offers the possibility to have physical implementations with other benefits in terms of energy-efficiency or processing complexity. For example, this restriction allows implementations where the first two stages of the WAX framework are performed in the analog domain using passive components, leading to important reductions in terms of energy consumption with respect to having them in baseband. On the other hand, the unitary restriction may also allow to simplify some of the computations required to compute the decentralized filters \cite{wax_journal,A_struct}, since these heavily rely on matrix inverses which would be trivial when enforcing the unitary constraint.\footnote{Note that the inverse of a unitary matrix is simply given by its conjugate transpose.} We will also show how to minimize the performance gap between the considered framework and a centralized processing approach.

\section{System model}
\begin{figure}[t]
	\centering
	\includegraphics[scale=0.51]{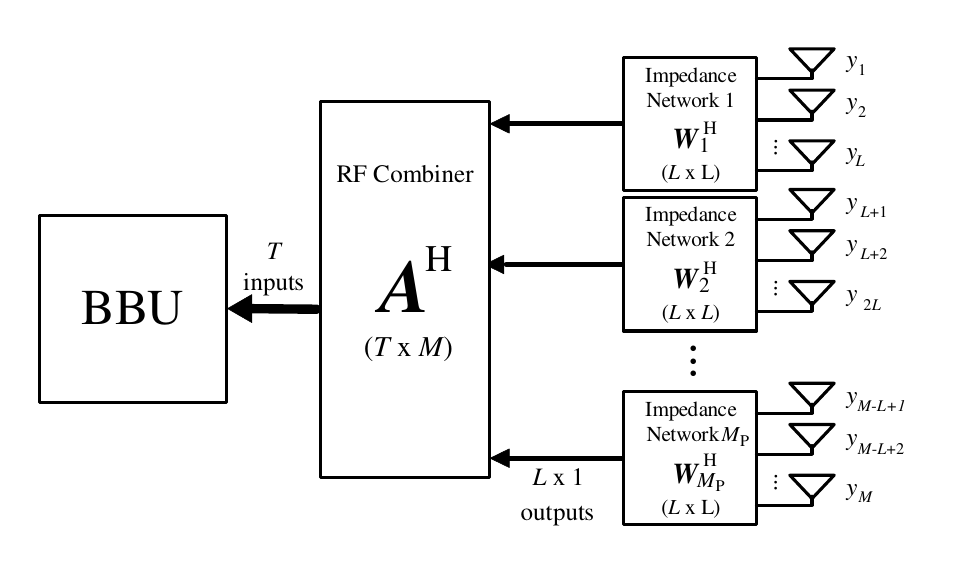}
    \vspace{-1em}
\caption{General decentralized framework under study.}
	\label{fig:sm}
 \vspace{-1em}
\end{figure}

We consider an uplink scenario where $K$ user equipments (UEs) communicate with an $M$-antenna BS, where $M\geq K$, through a narrowband channel. The complex-baseband equivalent received vector may be expressed as
\begin{equation}\label{eq:mimo}
    \bs{y} = \bs{H} \bs{s} + \bs{n},
\end{equation}
where $\bs{s}$ is the $K\times 1$ vector of symbols transmitted by each UE, modeled by $\bs{s}\sim\mathcal{CN}(\bs{0}_{K\times 1}, E_\mrm{s}\mathbf{I}_K)$, which corresponds to the capacity-achieving distribution \cite{heath2018foundations}, $\bs{H}$ is the $M\times K$ channel matrix known by the receiver, and $\bs{n}\sim\mathcal{CN}(\bs{0}_{M\times 1},N_0 \mathbf{I}_M)$ is the additive white gaussian noise (AWGN). We further assume that $\bs{H}$ is a \textit{randomly chosen} matrix \cite{A_struct}, i.e., any submatrix of it is assumed to be full-rank.

The processing is performed according to the WAX framework \cite{wax_journal}, leading to the following post-processed vector
\begin{equation}\label{eq:z}
    \bs{z} = \bs{X}^\mrm{H}\bs{A}^\mrm{H}\bs{W}^\mrm{H}\bs{y},
\end{equation}
where $\bs{W}=\diag(\bs{W}_1,\dots,\bs{W}_{\MP})$ is a $M\times M$ block diagonal matrix with the $L\times  L$ blocks $\{\boldsymbol{W}_m\}_{m=1}^{\MP}$ corresponding to the reconfigurable decentralized processing filters ($L\leq K$), $\bs{A}$ is the $M\times T$ matrix corresponding the fixed combining module ($T\leq M$), and $\bs{X}^\mrm{H}$ is a $T \times K$ matrix that considers the subsequent linear processing that may be applied at the CPU ($T\geq K$). 

In this work we consider (semi-)unitary restrictions on the decentralized filters $\{\boldsymbol{W}_m\}_{m=1}^{\MP}$, as well as on the combining module $\bs{A}$---i.e., $\bs{W}_m^\mrm{H}\bs{W}_m=\mathbf{I}_L$ and $\bs{A}^\mrm{H}\bs{A}=\mathbf{I}_T$. This design choice is intrinsically desirable to relieve complexity issues with matrix inversion, e.g., when computing each $\bs{W}_m$, and to get rid of impractical solutions with high condition numbers. However, another important benefit is that they allow for energy-efficient implementations as the one described in Fig.~\ref{fig:sm}, since the equivalent response of a lossless (purely reactive) impedance network has unitary restriction \cite{pozar}. 

\subsection{Background}
In \cite{wax_journal}, the information-lossless trade-off between $L$ and $T$, which respectively relate to the decentralized processing complexity and the interconnection bandwidth to the CPU, was characterized for unrestricted $\{\boldsymbol{W}_m\}_{m=1}^{\MP}$ and $\bs{A}$. The ability to perform information-lossless processing---i.e., such that $I(\bs{z};\bs{s})=I(\bs{y};\bs{s})$---was proved to be equivalent to the ability to perform the so called WAX decomposition of the channel, given by
\begin{equation}\label{eq:WAX}
    \bs{H} = \bs{W}\bs{A}\bs{X},
\end{equation}
where $\bs{W}$, $\bs{A}$, and $\bs{X}$ have direct connection to the processing from \eqref{eq:z}. Moreover, the availability of such decomposition determines the fundamental information lossless trade-off between decentralized processing complexity and interconnection bandwidth, given in \cite{wax_journal} as
\begin{equation}\label{eq:WAX_cond}
    T>\max\left(M\frac{K-L}{K}, K-1 \right)
\end{equation}
On other hand, in the regime of the trade-off where information-lossless processing is not available, a subotimal approach was proposed by solving the minimization problem
\begin{equation}\label{eq:app_mf}
\begin{aligned}
    \min_{\boldsymbol{X},\{\boldsymbol{W}_m\}_{m=1}^{\MP}} & \; \Vert \boldsymbol{A} \boldsymbol{X} - \boldsymbol{W}\boldsymbol{H} \Vert^2_{\mathrm{F}}, \\
   \mathrm{s.t.} \;\;\;
& \Vert \boldsymbol{X} \Vert_\mrm{F} + \Vert \boldsymbol{W} \Vert_\mrm{F} = c,
\end{aligned}
\end{equation}
which has closed-form solution by leveraging the eigenvalue decomposition \cite{wax_journal}.
\subsection{Problem formulation}
Given the new restrictions on $\{\boldsymbol{W}_m\}_{m=1}^{\MP}$, it is unclear if it is possible to perfectly fulfill \eqref{eq:WAX} for general $\bs{H}$---or equivalently to achieve information-lossless processing.\footnote{The restrictions on $\bs{A}$ have no real impact since it may still be assumed \textit{randomly chosen} within this constraint set, leading to the results from \cite{wax_journal}.} The general goal would then be to minimize the information loss, leading to the following optimization problem
\begin{equation}\label{eq:rate_max}
\begin{aligned}
    \max_{\{\boldsymbol{W}_m\}_{m=1}^{\MP}} &I(\bs{z};\bs{s})\\
    \mrm{s.t. }\;\;\; &\bs{W}_m^\mrm{H}\bs{W}_m = \mathbf{I}_L, \; \forall m
\end{aligned}
\end{equation}
where we should remark that $\bs{A}$ is assumed fixed in \eqref{eq:z} such that it fulfills the semi-unitary constraint. Note that we may disregard the processing $\bs{X}$ performed at the CPU---associated to the baseband unit (BBU) in Fig.~\ref{fig:sm}---since it can be trivially selected to fulfill $I(\bs{z};\bs{s})=I(\bs{A}^\mrm{H}\bs{W}^\mrm{H}\bs{y};\bs{s})$ \cite{wax_journal,A_struct}. 
\if Thus, the objective function in \eqref{eq:rate_max} may be substituted by \cite{mimo}
\begin{equation}
I(\bs{A}^\mrm{H}\bs{W}^\mrm{H}\bs{y};\bs{s})=\log \Big( \det\big( \mathbf{I}_T +\rho \bs{A}^\mrm{H}\bs{W}^\mrm{H}\bs{H}\bs{H}^\mrm{H}\bs{W}\bs{A}\big) \Big),
\end{equation}
where we have leveraged the unitary properties of $\bs{A}$ and $\bs{W}$, whose product is thus semi-unitary. \fi
An important observation is that, from the data-processing inequality \cite{inf_th}, $I(\bs{A}^\mrm{H}\bs{W}^\mrm{H}\bs{y};\bs{s})\leq I(\bs{y};\bs{s})$. Thus, having $\{\boldsymbol{W}_m\}_{m=1}^{\MP}$ such that $I(\bs{A}^\mrm{H}\bs{W}^\mrm{H}\bs{y};\bs{s})= I(\bs{y};\bs{s})$, directly gives a solution to \eqref{eq:rate_max}, associated to having information-lossless processing in the considered framework.

\section{Information loss minimization}
The problem defined in \eqref{eq:rate_max} is non-convex, and obtaining effective solutions with reasonable complexity seems highly non-trivial. We next propose to solve an alternative problem that would also give a solution to \eqref{eq:rate_max} under specific settings, while it gives approximate solutions for the general case.

\subsection{Information-lossless semi-unitary transformations}
Given the (semi-)unitary restriction on $\bs{A}$ and $\{\boldsymbol{W}_m\}_{m=1}^{\MP}$, it can be easily verified that the dimension reduction associated to the product $\bs{F}=\bs{W}\bs{A}$ corresponds to a semi-unitary matrix of dimension $M\times T$, i.e., $\bs{F}\in \mathcal{S}(M,T)$.\footnote{The notation $\mathcal{S}(M,T)$ is used to denote the Stiefel manifold, whose elements $\bs{F}\in\mathcal{S}(M,T)$ correspond to $M\times T$ matrices ($T\leq M$) fulfilling the semi-unitary constraint $\bs{F}^\mrm{H}\bs{F}=\mathbf{I}_T$ \cite{comm_grass}.} An important goal of this work is to particularize the information-lossless trade-off presented in \cite{wax_journal} to its counterpart, where $\bs{A}$ and $\bs{W}$ have (semi-)unitary constraints. The following lemma characterizes the set of information-lossless semi-unitary transformations in this context.

\begin{thm}\label{thm:inf_l}
    Let $\bs{F}_\mrm{L}\in \mathcal{S}(M,T)$ be an arbitrary semi-unitary information-lossless $M\times T$ transformation, i.e., such that $I(\bs{F}_\mrm{L}^\mrm{H}\bs{y};\bs{s})=I(\bs{y};\bs{s})$ for the signal model \eqref{eq:mimo} with $K\leq T\leq M$. We can then find two unitary matrices $\bs{Q}\in \mathcal{U}(T)$, and $\bs{Q}_0\in \mathcal{U}(M-K)$ such that
    \begin{equation}\label{eq:F_L}
        \bs{F}_\mrm{L}(\bs{Q},\bs{Q}_0) = \begin{bmatrix}\widetilde{\bs{U}}_{\bs{H}} &
    \bs{N}_{\bs{H}}\bs{Q}_0 \end{bmatrix}\begin{bmatrix}\bs{Q}\\
    \bs{0}
    \end{bmatrix},
    \end{equation}
where $\widetilde{\bs{U}}_{\bs{H}}\in \mathcal{S}(M,K)$ is a given semi-unitary matrix defining the signal space of $\bs{H}$, and $\bs{N}_{\bs{H}}\in \mathcal{S}(M,(M-K))$ is a given semi-unitary matrix defining the null-space of $\bs{H}$. Note that $\widetilde{\bs{U}}_{\bs{H}}$ and $\bs{N}_{\bs{H}}$ are mutually orthogonal, and they can be uniquely identified with the respective submatrices of the unitary matrix obtained from a specific implementation of the QR-decomposition of $\bs{H}$.
\begin{skproof}
The proof consists of comparing the mutual information expressions for $I(\bs{y};\bs{s})$ and $I(\bs{F}_\mrm{L}^\mrm{H}\bs{y};\bs{s})$ with the channel matrix given as a QR-decomposition
\begin{equation}
    \bs{H}=\begin{bmatrix}
        \widetilde{\bs{U}}_{\bs{H}} &
        \bs{N}_{\bs{H}}
    \end{bmatrix} \begin{bmatrix}
        \widetilde{\bs{R}}_{\bs{H}} \\
        \bs{0}
    \end{bmatrix}.
\end{equation}
The condition for $\bs{F}_\mrm{L}$ to be information-lossless can be then written as
\begin{equation}
\widetilde{\bs{U}}_{\bs{H}}\bs{F}_\mrm{L}\bs{F}_\mrm{L}^\mrm{H}\widetilde{\bs{U}}_{\bs{H}}^\mrm{H}=\mathbf{I}_K,
\end{equation}
which after some matrix manipulation leads to \eqref{eq:F_L}. An intuitive argument for \eqref{eq:F_L} is that an information-lossless semi-unitary transformation should consist of an arbitrary unitary combination of the $K$-dimensional signal space, and a $(T-K)$-dimensional semi-unitary combination of the null-space. A detailed proof may be included in the extended version.
\if
Consider a fixed QR-decomposition of the channel matrix expressed as
\begin{equation}
    \bs{H}=\begin{bmatrix}
        \widetilde{\bs{U}}_{\bs{H}} &
        \bs{N}_{\bs{H}}
    \end{bmatrix} \begin{bmatrix}
        \widetilde{\bs{R}}_{\bs{H}} \\
        \bs{0}
    \end{bmatrix},
\end{equation}
where, since $\bs{H}$ is randomly chosen and thus full-rank, $\tilde{\bs{R}}_{\bs{H}}$ is a full-rank $K\times  K$ upper triangular matrix, $\widetilde{\bs{U}}_{\bs{H}}\in \mathcal{S}(M\times K)$ spans the signal space of $\bs{H}$, and $\widetilde{\bs{N}}_{\bs{H}}\in \mathcal{S}(M\times M-K)$ spans its null-space. Given $\bs{F}\in \mathcal{S}(M,T)$, we have that \cite{mimo}
    \begin{equation}\label{eq:F_inf}
    \begin{aligned}
    I(\bs{F}^\mrm{H}\bs{y};\bs{s})&=\log \Big( \det\big( \mathbf{I}_T + \bs{F}^\mrm{H}\bs{H}\bs{H}^\mrm{H}\bs{F}\big) \Big)\\
    &=\log \left( \det\big(\bs{F}^\mrm{H} \bs{Q}_{\bs{H}} \left( \mathbf{I}_K + \begin{bmatrix}
        \bs{\Theta} & \bs{0} \\
        \bs{0} & \bs{0}
\end{bmatrix}\right)\bs{Q}_{\bs{H}}^\mrm{H}\bs{F} \right)
    \end{aligned}
    \end{equation}
    where $\bs{\Theta}=\widetilde{\bs{R}}_{\bs{H}}\widetilde{\bs{R}}_{\bs{H}}^\mrm{H}$. Note that we have ignored without loss of generality the ratio $E_s/N_0$, which can be absorbed by \bs{H}. In order for $\bs{F}$ to be an information-lossless transformation, we need to have $I(\bs{F}^\mrm{H}\bs{y};\bs{s})=I(\bs{y};\bs{s})$, where we have
    \begin{equation}
    \begin{aligned}
        I(\bs{y};\bs{s})&=\log \Big( \det\big( \mathbf{I}_M + \bs{H}\bs{H}^\mrm{H}\big) \Big)\\
        &=\log \Big( \det\big( \mathbf{I}_K + \widetilde{\bs{R}}_{\bs{H}}\big) \Big),
    \end{aligned}
    \end{equation}
    where $\widetilde{\bs{\Lambda}}_{\bs{H}}$ is the $K\times K$ diagonal matrix of non-zero eigenvalues of $\bs{H}^\mrm{H}\bs{H}$.
\fi
\end{skproof}
\end{thm}

From Theorem~\ref{thm:inf_l}, and given the semi-unitary nature of the product $\bs{F}=\bs{A}\bs{W}$, if information-lossless processing was available, we should then be able to find $\boldsymbol{Q}\in \mathcal{U}(T)$ and $\boldsymbol{Q}_0\in \mathcal{U}(M-K)$ such that
\begin{equation}\label{eq:inf_l_unit}
    \bs{W} \bs{A} =  \bs{F}_\mrm{L}(\bs{Q},\bs{Q}_0).
\end{equation}
Alternatively, we can think of $\bs{Q}$ and $\bs{Q}_0$ as the arbitrary $T\times T$ and $(M-K)\times (M-K)$ unitary matrices that capture the available degrees of freedom (DoFs) for achieving information-lossless semi-unitary processing. 

An interesting alternative to solving \eqref{eq:rate_max} is to minimize the information-loss by minimizing the distance between the achievable semi-unitary processing in the considered framework, and an arbitrary information-lossless semi-unitary transformation. We thus propose to minimize the distance between $\bs{W}\bs{A}$ and $\bs{F}(\bs{Q},\bs{Q}_0)$, which may be conveniently measured in terms of the Frobenius norm $\Vert\bs{W}\bs{A}-\bs{F}(\bs{Q},\bs{Q}_0)\Vert^2$. The available DoFs are then captured in the matrices $\bs{Q}$, $\bs{Q}_0$, and $\{\boldsymbol{W}_m\}_{m=1}^{\MP}$, all of which are subject to unitary constraints. This leads to the following optimization problem
\begin{equation}\label{eq:min_inf-loss}
\begin{aligned}
    \min_{\{\boldsymbol{W}_m\}_{m=1}^{\MP},\bs{Q},\bs{Q}_0}&D_\mrm{L}(\bs{W},\bs{Q},\bs{Q}_0) \triangleq \Vert \bs{W} \bs{A}-\bs{F}(\bs{Q},\bs{Q}_0)\Vert^2_\mrm{F}\\
    \mrm{s.t. \;\;}& \bs{W}_m^\mrm{H}\bs{W}_m=\mathbf{I}_L, \; \forall m,\\
    &  \bs{Q}^\mrm{H}\bs{Q}=\mathbf{I}_T\\
    & \bs{Q}_0^\mrm{H}\bs{Q}_0=\mathbf{I}_{M-K}  .
\end{aligned}
\end{equation}
It is important to remark that, if information-lossless processing is available in the considered framework, the set $\{\boldsymbol{W}_m\}_{m=1}^{\MP}$, obtained from solving \eqref{eq:min_inf-loss}, would also correspond to a solution for \eqref{eq:rate_max}. On the other hand, information-lossless processing is achievable if and only if a solution to \eqref{eq:min_inf-loss} is available with $\Vert \bs{W} \bs{A}-\widetilde{\bs{U}}_{\bs{H}}\bs{Q}\Vert^2=0$. In general, \eqref{eq:min_inf-loss} would still intuitively give an approximate solution to \eqref{eq:rate_max}. Note the similarity to the approach based on \eqref{eq:app_mf} for the unconstrained case, which also leads to information-lossless processing whenever available. However, said approach relies on the solution $\{\boldsymbol{W}_m\}_{m=1}^{\MP}$ consisting of full-rank matrices, while the current method enforcing unitary constraints is more robust towards cases where this may not naturally happen, e.g., for poorly constructed $\bs{A}$ or null blocks in $\bs{H}$ \cite{wax_journal,cell-free_wax}.

\subsection{Information-lossless processing distance minimization}
Next, we study an approach to tackle the minimization problem defined in \eqref{eq:min_inf-loss}. Considering that both $\bs{W}\bs{A}$ and $\bs{F}_\mrm{L}(\bs{Q},\bs{Q}_0)$ give semi-unitary matrices, we can express the objective function as
\begin{equation}\label{eq:f_obj}
\begin{aligned}
    D_\mrm{L}(\bs{W},\bs{Q},\bs{Q}_0) = 2T-2\Re\{\mrm{tr}(\bs{A}^\mrm{H}\bs{W}^\mrm{H}\bs{F}_\mrm{L}(\bs{Q},\bs{Q}_0))\}.
\end{aligned}
\end{equation}
We can then reformulate the optimization problem in \eqref{eq:min_inf-loss} as
\begin{equation}\label{eq:max_tr}
\begin{aligned}
    \max_{\{\boldsymbol{W}_m\}_{m=1}^{\MP},\bs{Q},\bs{Q}_0}& \mathcal{J}(\bs{W},\bs{Q},\bs{Q}_0) \triangleq \Re\{\mrm{tr}(\bs{A}^\mrm{H}\bs{W}^\mrm{H}\bs{F}_\mrm{L}(\bs{Q},\bs{Q}_0))\}\\
    \mrm{s.t. \;\;}& \bs{W}_m^\mrm{H}\bs{W}_m=\mathbf{I}_L, \; \forall m,\\
    &  \bs{Q}^\mrm{H}\bs{Q}=\mathbf{I}_T\\
    & \bs{Q}_0^\mrm{H}\bs{Q}_0=\mathbf{I}_{M-K}  .
\end{aligned}
\end{equation}
If we substitute \eqref{eq:F_L} in \eqref{eq:max_tr}, it may seem that relation between the unitary optimization variables in \eqref{eq:max_tr} is too complicated to attempt a direct solution. However, if we consider each variable individually, i.e., $\{\boldsymbol{W}_m\}_{m=1}^{\MP}$,$\bs{Q}$, and $\bs{Q}_0$, we may note that each of them have only linear effect on the objective function $\mathcal{J}(\bs{W},\bs{Q},\bs{Q}_0)$. The following lemma gives a closed-form solution to problems of the form \eqref{eq:max_tr} over a unitary matrix variable with linear effect on the objective function.

\begin{lem}\label{lem:opt_lin}
Consider the optimization problem
\begin{equation}\label{eq:max_tr_lem}
\begin{aligned}
    \max_{\bs{U}}&\;\; \Re\{\mrm{tr}(\bs{U}^\mrm{H}\bs{B}) \}\\
    \mrm{s.t. \;\;}& \bs{U}\in \mathcal{U}(N),
\end{aligned}
\end{equation}
where $\bs{B}$ may be an arbitrary $N\times N$ matrix with singular value decomposition (SVD) $\bs{B}=\bs{U}_{\bs{B}}\bs{S}_{\bs{B}}\bs{V}_{\bs{B}}^\mrm{H}$. The solution to \eqref{eq:max_tr_lem} is given by
\begin{equation}\label{eq:sol_max}
    \bs{U}_\mrm{opt} = \bs{U}_{\bs{B}}^\mrm{H}\bs{V}_{\bs{B}}.
\end{equation}
\begin{skproof}
    A formal proof, which may be included in the extended version, can be obtained by defining the Riemannian gradient over the unitary group and finding the stationary point where it vanishes. However, an intuitive argument is that the solution to \eqref{eq:max_tr_lem} is obtained by positively combining the singular values of $\bs{B}$, leading to \eqref{eq:sol_max}.
\end{skproof}
\end{lem}

Using Lemma~\ref{lem:opt_lin} we can then solve \eqref{eq:max_tr} by iteratively maximizing the objective function over $\{\boldsymbol{W}_m\}_{m=1}^{\MP}$,$\bs{Q}$, and $\bs{Q}_0$, where we can get a close-form maximum for each step. Note that $\mathcal{J}(\bs{W},\bs{Q},\bs{Q}_0)$ is upper-bounded by $T$ since $D_\mrm{L}(\{\bs{W},\bs{Q},\bs{Q}_0)$, given in \eqref{eq:f_obj}, is a distance metric lower-bounded by $0$. Hence, the iterative maximization of $\mathcal{J}(\bs{W},\bs{Q},\bs{Q}_0)$ converges to the global maximum for \eqref{eq:max_tr}, and incidentally to the global minimum for \eqref{eq:min_inf-loss}. On the other hand, if we get a solution that attains $\mathcal{J}(\bs{W},\bs{Q},\bs{Q}_0)=T$ we can also conclude that the resulting processing is information lossless. The only complication left is to express $\mathcal{J}(\bs{W},\bs{Q},\bs{Q}_0)$ in the form considered in Lemma~\ref{eq:max_tr_lem} for each unitary variable, as shown next.

We may start by expressing $\mathcal{J}(\bs{W},\bs{Q},\bs{Q}_0)$ as
\begin{equation}
    \mathcal{J}(\bs{W},\bs{Q},\bs{Q}_0)=\sum_{m=1}^{\MP}\Re\{\mrm{tr}\big(\bs{W}^\mrm{H}_{m}\bs{F}_\mrm{L}(\bs{Q},\bs{Q}_0)\bs{A}_m^H\big)\},
\end{equation}
where $\bs{A}_m$ corresponds to the $m$th $L\times T$ row block of $\bs{A}=[\bs{A}_1^\mrm{T},\dots,\bs{A}_{\MP}^\mrm{T}]^\mrm{T}$. We can then define
\begin{equation}
    \bs{B}_{\bs{W}_m}(\bs{Q},\bs{Q}_0)=\bs{F}_\mrm{L}(\bs{Q},\bs{Q}_0)\bs{A}_m^H,
\end{equation}
whose SVD gives the solution over $\bs{W}_m$ as \eqref{eq:sol_max}
\begin{equation}
    \bs{W}_{m,\mrm{opt}}(\bs{Q},\bs{Q}_0)=\bs{U}_{\bs{B}_{\bs{W}_m}}(\bs{Q},\bs{Q}_0)\bs{V}_{\bs{B}_{\bs{W}_m}}^\mrm{H}(\bs{Q},\bs{Q}_0).
\end{equation}
If we substitute \eqref{eq:F_L} in \eqref{eq:max_tr}, we can also express $\mathcal{J}(\bs{W},\bs{Q},\bs{Q}_0)$ as
\begin{equation}\label{eq:obj_F_L}
    \mathcal{J}(\bs{W},\bs{Q},\bs{Q}_0)=\Re\left\{\mrm{tr}\left(\bs{Q}^\mrm{H} \begin{bmatrix}\widetilde{\bs{U}}_{\bs{H}}^\mrm{H}\\
    [\bs{Q}_0]_{:,1:T-K}^\mrm{H} \bs{N}_{\bs{H}}^\mrm{H}\end{bmatrix}\bs{W}\bs{A}\right)\right\},
\end{equation}
where $[\bs{Q}_0]_{:,1:T-K}$ is the matrix formed by the first $T-K$ columns of $\bs{Q}_\mrm{0}$. We can then define
\begin{equation}
    \bs{B}_{\bs{Q}}(\bs{W},\bs{Q}_0)=\begin{bmatrix}\widetilde{\bs{U}}_{\bs{H}}^\mrm{H}\\
    [\bs{Q}_0]_{:,1:T-K}^\mrm{H} \bs{N}_{\bs{H}}^\mrm{H}\end{bmatrix}\bs{W}\bs{A},
\end{equation}
whose SVD gives the solution over $\bs{Q}$ as
\begin{equation}
    \bs{Q}_{\mrm{opt}}(\bs{W},\bs{Q}_0)=\bs{U}_{\bs{B}_{\bs{Q}}}(\bs{W},\bs{Q}_0)\bs{V}_{\bs{B}_{\bs{Q}}}^\mrm{H}(\bs{W},\bs{Q}_0).
\end{equation}
Finally, we can rewrite \eqref{eq:obj_F_L} as
\begin{equation}
\begin{aligned}
    \mathcal{J}(\bs{W},\bs{Q},\bs{Q}_0)=&\Re\big\{\mrm{tr}\big(\bs{Q}_0^\mrm{H}\bs{N}_{\bs{H}}^\mrm{H}\bs{W}\bs{A}\big[[\bs{Q}]_{K+1:T,:}^\mrm{H} \;\; \bs{0}\big]\big)\big\}\\
    &+\Re\{\mrm{tr}(\widetilde{\bs{U}}_{\bs{H}}^\mrm{H}\bs{W}\bs{A}[\bs{Q}]_{1:K,:}^\mrm{H})\},
\end{aligned}
\end{equation}
where the second term has no dependency on $\bs{Q}_0$ so we can ignore it when optimizing over it. We can then define
\begin{equation}
    \bs{B}_{\bs{Q}_0}(\bs{W},\bs{Q})=\bs{N}_{\bs{H}}^\mrm{H}\bs{W}\bs{A}\big[[\bs{Q}]_{K+1:T,:}^\mrm{H} \;\; \bs{0}\big],
\end{equation}
whose SVD gives the solution over $\bs{Q}_0$
\begin{equation}
    \bs{Q}_{0,\mrm{opt}}(\bs{W},\bs{Q})=\bs{U}_{\bs{B}_{\bs{Q}_0}}(\bs{W},\bs{Q})\bs{V}_{\bs{B}_{\bs{Q}_0}}^\mrm{H}(\bs{W},\bs{Q}).
\end{equation}

Altogether, we have derived the expressions to solve \eqref{eq:max_tr}, and thus \eqref{eq:min_inf-loss}, by iteratively optimizing in closed-form over each unitary variable, $\{\boldsymbol{W}_m\}_{m=1}^{\MP}$,$\bs{Q}$, and $\bs{Q}_0$. The proposed approach employs said iterative optimization until convergence, e.g., when the the objective function gives the same values up to threshold after updating the optimization variables. We next consider a baseline approach to assess the performance of our proposed method.

\subsection{Baseline approach}
Previous work dealing with the WAX framework \cite{wax_journal} includes a closed-form approach to solve the suboptimal information-loss minimization problem proposed in \eqref{eq:app_mf}, which also leads to information-lossless processing when the condition \eqref{eq:WAX_cond} is fulfilled.
We may thus consider a baseline approach by simply projecting onto the unitary group the set of decentralized filters $\{\boldsymbol{W}_m\}_{m=1}^{\MP}$ obtained from solving \eqref{eq:app_mf} (with fixed semi-unitary combining module $\bs{A}$). When \eqref{eq:WAX_cond} is fulfilled, this approach corresponds to projecting onto the unitary group the $\{\boldsymbol{W}_m\}_{m=1}^{\MP}$ obtained from the WAX decomposition of the channel matrix \eqref{eq:WAX}. Moreover, the projection of an arbitrary full-rank matrix onto the unitary group, in terms of Frobenius and spectral norm, is given by the polar factor from the polar decomposition of said matrix \cite{polar_proj}. Thus, our baseline approach will consist on defining $\{\boldsymbol{W}_m\}_{m=1}^{\MP}$ as the polar factors of the respective matrices obtained when solving \eqref{eq:app_mf}.

\section{Numerical Results}
In order to assess the performance of the considered methods, we define the capacity ratio as the ratio between the channel capacity, given by $I(\bs{y};\bs{s})$ for the signal model \eqref{eq:mimo}, and the mutual information achieved after the respective decentralized processing, given by $I(\bs{A}^\mrm{H}\bs{W}^\mrm{H}\bs{y};\bs{s})$ in the considered framework. In the case of information-lossless processing, the full channel capacity will be available, leading to a capacity ratio of $100$\%. In the original unconstrained WAX framework this is  achieved whenever \eqref{eq:WAX_cond} is fulfilled, so we can obtain the minimum $T$ for information-lossless processing in that case as
\begin{equation}\label{eq:Tmin}
    T_\mrm{min}=\max\left(K,\left\lfloor \frac{M(K-L)}{K+1} \right\rfloor\right).
\end{equation}

\begin{figure}
  \begin{subfigure}{\columnwidth}
    \includegraphics[width=\linewidth]{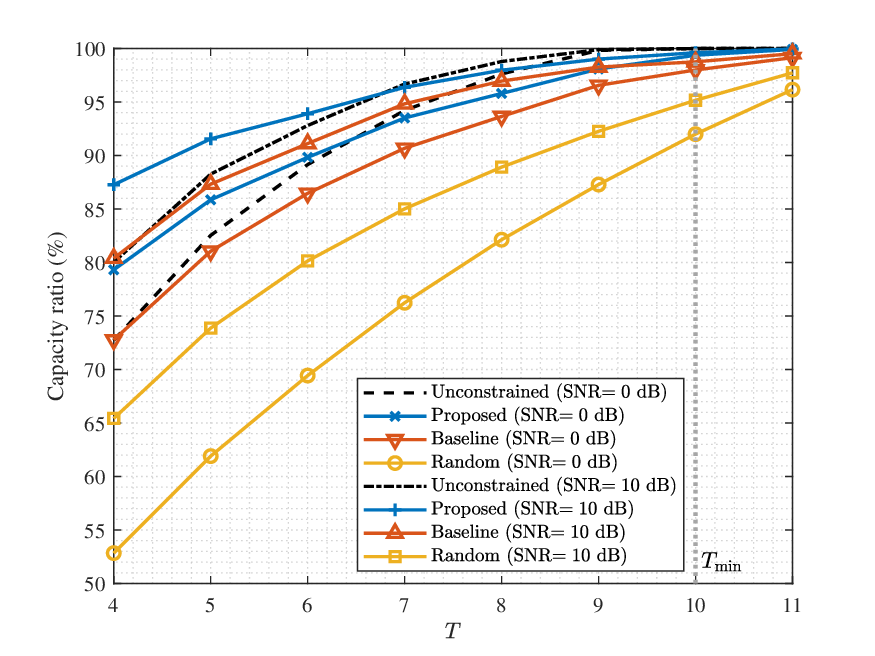}
    \caption{$L=1$}
    \label{fig:inf_loss_L1}
  \end{subfigure}
  \hfill 
  \begin{subfigure}{\columnwidth}
    \includegraphics[width=\linewidth]{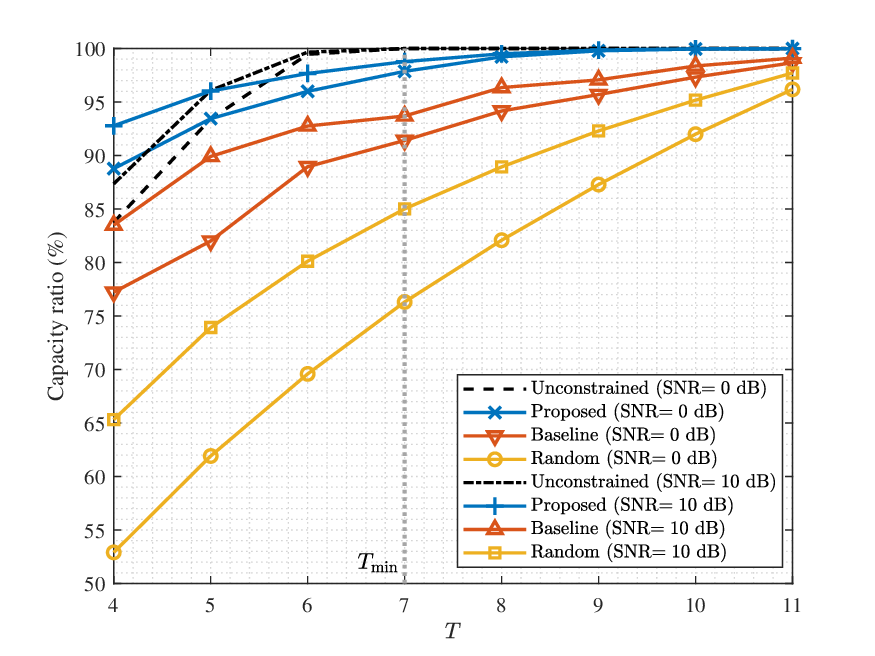}
    \caption{$L=2$}
    \label{fig:inf_loss_L2}
  \end{subfigure}
  \hfill 
  \begin{subfigure}{\columnwidth}
    \includegraphics[width=\linewidth]{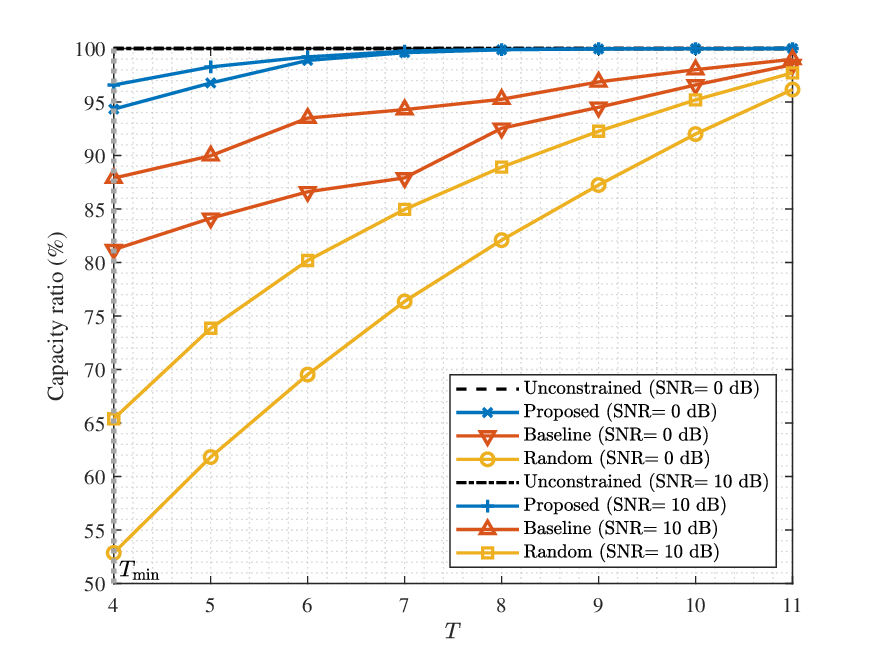}
    \caption{$L=3$}
    \label{fig:inf_loss_L3}
  \end{subfigure}
  \caption{Capacity ratio for $M=12$ and $K=4$.}
    \label{fig:rates}
\end{figure}

Fig.~\ref{fig:rates} plots the capacity ratio versus the interconnection bandwidth (given by $T$) averaged over $10^4$ realizations of an IID Rayleigh fading channel with normalized unit power per entry \cite{mimo}. We have considered a system with $M=12$ antennas and $K=4$ UEs, and included the results for different values of decentralized processing complexity, $L\in\{1,2,3\}$. Note the correspondence between the case with $L=1$ and a common hybrid beamforming scheme, where each antenna would apply a phase shift before combining the outputs in the analog domain. Apart from the results of the proposed approach, we have included results for the baseline approach previously discussed, as well as for the unconstrained approach, which also solves \eqref{eq:app_mf}, but skips the projection step, i.e., leading to information-lossless processing for $T\geq T_\mrm{min}$. To understand the available gain when using elaborate approaches, we have also included results for random isotropic selection of $\{\boldsymbol{W}_m\}_{m=1}^{\MP}$ with unitary constraints (note that these still lead information-lossless processing for $T=M$). The plots include results for a low and a high SNR scenario. However, we should note that, when information-losslesss processing is available, the capacity ratio should be $100$\% independently of the SNR.\footnote{A direct consecuence of Theorem~\ref{thm:inf_l} is that the availability of information-lossless processing is independent of the SNR.} In all cases, the proposed approach outperforms the baseline approach, with a wider margin for lower values of $T$. Moreover, we also get a greater margin for larger values of $L$, hinting that our proposed approach is more effective at exploiting the decentralized processing resources. For low enough values of $T$, we can even see some gain with respect to the unconstrained approach, which means that, even though both approaches are suboptimal with respect to the original problem \eqref{eq:rate_max} (or, respectively, its unconstrained version) when information-lossless processing is not available, the proposed approach has lower performance gap with respect to the optimal solution for \eqref{eq:rate_max}. Finally, the proposed approach seems to reach information lossless processing for large enough values of $T$, e.g., $T\geq 10$ for $L=2$ and $T\geq 9$ for $L=3$. This fact indicates the existence of a degraded version of the trade-off \eqref{eq:WAX_cond} from \cite{wax_journal} when considering unitary restrictions in the decentralized processing. Future work may explore a deeper characterization of said degraded trade-off.

\section{Conclusion}
We have studied a general framework for decentralized architectures with unitary constraints on the decentralized processing. These constraints allow considering the trade-off between interconnection bandwidth and decentralized processing complexity with passive analog processing schemes. We have characterized the structure of an arbitrary information-lossless semi-unitary transformation, and used it to propose an approach for finding the decentralized processing filters minimizing the information loss. The numerical results show the potential of the proposed approach, which seems to even achieve information-lossless processing under certain parameter settings. Future work may further explore the degraded information-lossless trade-off for decentralized architectures with unitary constraints, which was characterized in previous literature for the unconstrained case.

\bibliographystyle{IEEEtran}
\balance
\bibliography{IEEEabrv,wax}

%
\IEEEpeerreviewmaketitle

\end{document}